\documentclass{llncs}
	\usepackage{amscd} %% commutative diagrams
	\usepackage{amssymb}
	\usepackage{url}
	\usepackage{graphicx}
	\usepackage{amsmath}
	\usepackage{stmaryrd}
	\usepackage{epsfig}
	\usepackage{subfigure}
	\usepackage{mdwlist}
	\usepackage{enumitem}
	\usepackage{color}
	\usepackage[boxed,ruled,noend,vlined,linesnumbered]{algorithm2e}
	\usepackage[noend]{algorithmic}
	\usepackage[usenames,dvipsnames]{xcolor}
	\usepackage[linesnumbered]{algorithm2e}
	\newtheorem{cor}[theorem]{Corollary}
	\newtheorem{lem}[theorem]{Lemma}
	\newtheorem{prop}[theorem]{Proposition}
	
	\newtheorem{defn}[theorem]{\textbf{Definition}}
	\newtheorem{exmp}[theorem]{\textbf{Example}}
	\newcommand{\MM}{\mathcal{M}}
	
	\newcommand{\too}{\twoheadrightarrow}
		 \usepackage[hmargin=4.21cm,vmargin=4.21cm]{geometry}

\begin{document}
	%%%
	%%
	\title{Web Maps and their Algebra}
 \author{Valeria Fionda \inst{1},  Claudio Gutierrez\inst{2}, Giuseppe Pirr\'o\inst{1}}
 \authorrunning{Valeria Fionda,  Claudio Gutierrez, Giuseppe Pirr\'o}
 %%
 %\vspace{-.2cm}
 %%
 \institute{KRDB, Free University of Bozen-Bolzano, Bolzano, Italy\\ 
 %\email{lastname@inf.unibz.it}\\
 \and DCC, Universidad de Chile, Santiago, Chile \\
  %\email{cgutierr@dcc.uchile.cl}\\
 }
	\maketitle
	%%%%% ABSTRACT%%%%%%%%%%%%%%%%%%%%%%
	
	%
	\vspace{-.6cm}
\begin{abstract}
A map is an abstract visual representation of a region, taken from a given space, usually designed for final human consumption. Traditional cartography focuses on the mapping of Euclidean spaces by using some distance metric. In this paper we aim at mapping the Web space by leveraging its relational nature. We introduce a general mathematical framework for maps and an algebra and discuss the feasibility of maps suitable for interpretation not only by humans but also by machines.
 \end{abstract}

\section{Introduction}
% % %
\vspace{-.2cm}
The Web is a virtually infinite space of interconnected resources. The common medium to access it is via navigation enabled by browsers. To cope with the size of this huge (cyber)space, Web users need to track, record and specify conceptual regions of information on the Web, for their own use, for exchanging, for further processing and, to avoid to incur in the ``lost in the cyberspace'' syndrome.
 %%% most popular competitors of the idea 
%
There are many tools (partially) addressing this need. The most traditional and popular are bookmarks: a list of URLs, sometimes categorized by tags. This idea has been enhanced to incorporate, for instance, social features (share, rank, tag bookmarks) and/or annotations of different types of data (e.g., not only URLs but also documents). Deliciuos and Diigo are among the most popular bookmarking systems.
	%%%
%
    Other approaches go beyond bookmarks and enable to organize URLs to also highlight connections between them. Results are grouped and presented on the form of a graphical map, which simulates the idea of a virtual map of a Web region. Some examples are search engines like Tag Galaxy, navigational history tools (e.g.,~\cite{Doemel1995}), visual HTML site maps (for users) and atlases of the Web (e.g.,~\cite{Dodge2001}). More recent approaches focus on providing visual representations of specific domains such as publications or news (e.g.,~\cite{Shahaf2012a}). 

 When it comes to the Web scale, these tools present drawbacks. First, they are designed for human visualization; hence they do not consider automatic processing, composition and reuse, thus hindering the automation of the process of creating, exchanging, combining and interpreting maps. Second, they do not include formal/provable connectivity relations between the URLs chosen; formal notions of quality, granularity and scope; and formal provable relations between the map and the region it represents, thus obstructing the generation of formal deductions from them.

\begin{figure}[tb]
	 \centering
	 \includegraphics[width=.65\columnwidth]{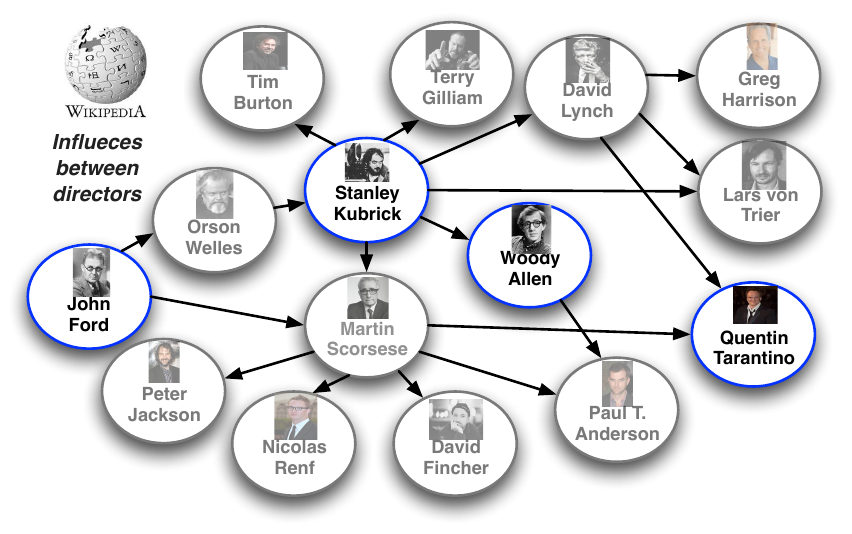}
	 \vspace{-.4cm}
	 \caption{A Web region taken from Wikipedia.}
	 \label{fig:region}
	 \vspace{-.6cm}
	 \end{figure}

\begin{figure}[!h]
	 \centering
	 \includegraphics[width=.85\columnwidth]{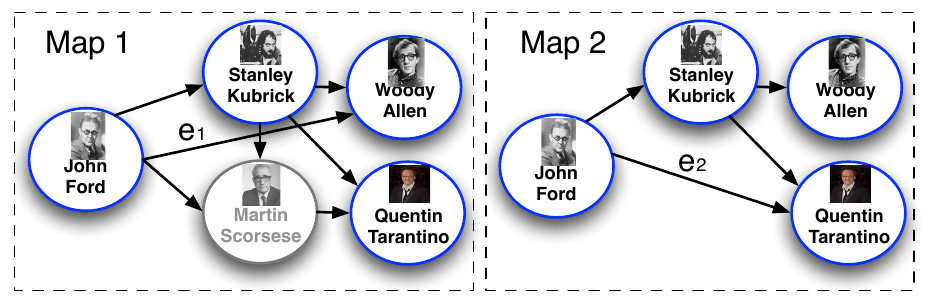}
	 	 \vspace{-.3cm}
	 \caption{Two possible maps of the region in Fig.~\ref{fig:region}.}
	 \label{fig:maps-comp}
	 \vspace{-.6cm}
	 \end{figure}

\smallskip
\noindent
\textbf{Regions and Maps.} Fig.~\ref{fig:region} shows a Web region taken from Wikipedia. In the region, the user \textit{Syd} has marked his favourite directors, that is, J. Ford, S. Kubrick, W. Allen and Q. Tarantino. The region besides these nodes also contains other nodes (lighter nodes). A question arises: how does a good map of \textit{Syd's} favourite directors look like?
	%%%
	%%
%
Fig.~\ref{fig:maps-comp} shows two possible maps for the region in Fig.~\ref{fig:region}. Map 1 contains more nodes (e.g., M. Scorsese) and edges (e.g., the edge $e_1$) than Map 2. This latter map adopts a specific conciseness strategy; it \textit{minimizes} the number of nodes and edges to keep connectivity among pairs of distinguished nodes. The node M. Scorsese is not included since it is not a distinguished node, but the connectivity between J. Ford and Q. Tarantino (both distinguished nodes) is still maintained via the direct edge $e_2$. The edge $e_1$ in Map 1 is not included in Map 2 because the connectivity between J. Ford and W. Allen is still maintained via S. Kubrick and there is no other path in the region going from J. Ford to W. Allen only passing for non distinguished nodes. 
\vspace{-.6cm}
	%%%
	%%
	
The idea of a map of a region is essentially that of \textit{reflecting} in a \textit{concise way} information in the region in terms of \textit{connectivity} among distinguished nodes (e.g., \textit{Syd's} favourite directors). However, how much of the original region has to be included in the map? The writer J. L. Borges scoffs at perfectly accurate maps when he talks about a ``map of the Empire.. which coincided point for point with it''. At the other extreme, \textit{minimal} maps are those that only include nodes with no information about their connectivity (e.g., bookmarks). In between there are maps that besides the distinguished nodes also provide information about their connectivity (e.g., Map 1 and Map 2). A flexible mapping framework should consider different types of maps.
% %
% %

\smallskip
\noindent
\textbf{Maps on the Web.} With the advent of the Web of Data~\cite{Heath2011}, maps to describe and navigate information on the Web in a machine-processable way become more feasible. The key new technical support are: i) the availability of a standard infrastructure, based on the Resource Description Framework (RDF), for the publishing and interlinking of structured data on the Web; ii) an active community of developers.
% % %
% % %
% % % %

\noindent
\textbf{Related Work.}
%
	%%%
	%%%
	Since the Web can be modeled as a graph, we review the general problem of graph summarization. In this respect, several approaches have been proposed (e.g., \cite{Faloutsos2004,Adibi2004,Zhou2010}) that address the following problem: given a graph structure $N$, determine a function $F$ in order to find a simplified structure $N^s$ satisfying some requirements. $F$ usually leverages some techniques such as data mining or information content and aims at simplifying the \textit{whole} graph structure.	% %
 There is a crucial difference between summaries, indexes and maps: a summary is a brief statement of the main points of something while an index is an alphabetical list with references to the place where some piece of information can be found. None of them give a well-founded and reusable ``representation'' of the object being summarized or indexed.
 	%%
 	%%%

 By moving our focus on the construction of maps, Dodge~\cite{Dodge2001} in is his book \textit{Atlas of the Cyberspace}, provides a comprehensive overview of visual representations of \textit{digital landscapes} on the Web. A recent information visualization paradigm used to summarize information is that of \textit{metro maps} (e.g.,~\cite{Shahaf2012a}). Other strands of research related to ours are (visual) navigational histories site maps and bookmarks \cite{van1989,Mceneaney1999,Doemel1995}. These approaches are designed for human usage and are mainly oriented to visualization not allowing automatic processing, composition and reuse. Moreover, they do not include formal/provable relations of connectivity between the URLs chosen; formal notions of quality, granularity and scope; and formal provable relations between the map and the region it represents; thus obstructing the generation of formal deductions from them.
% %
%%

\smallskip
\noindent
\textbf{Contributions.}
 In this paper we develop the theoretical basis and present a procedure to deal with the formal notion of map of the Web. We leverage formal techniques from graph theory and instantiate our proposal on the semantic infrastructure given by Linked Data~\cite{bizer09a}, Semantic Web tools and languages today available. Our final aim is to enable the creation of maps of Web regions that are machine-processable, endowed with provable formal properties, reusable, that can be composed, and of course, feasible to be constructed.
There are several challenges toward the developing of Web maps. First, given a region of the Web (a directed graph with suitable metadata on its nodes and edges), provide a definition of map of it with desirable formal properties. Second, given a user need or a conceptual notion, enable the specification of a region of the Web that represents or encompasses it. Third, devise algorithms and compose the procedures efficiently.
%
 %For sake of space in this paper we do report all the proofs and algortimhs. The interested reader can consult an extented version of the present paper~\cite{full_paper}.

\medskip
\noindent
\textbf{Roadmap.}
 In Section~\ref{sec:maps} we introduce a formal framework to cope with the notion of map as a means to abstract a region of a graph and formally study the properties of different types of maps. We also present an algebra for maps and efficient algorithms to compute maps. In Section~\ref{sec:framework} we briefly discuss how to apply our framework to the Web. Finally, Section~\ref{sec:conclusions} draws some conclusions and sketches future work.
%%%%	
%%%%
	%%%%
	%%%%
\section{Formal Maps on the Web}
\label{sec:maps}
\vspace{-.3cm}
The study of making maps is know as cartography~\cite{Robinson1995}. Cartography relies on the human mind's ability to read complex information represented in the map.
	In the following we provide a formal and general definition of map of a graph where nodes represent objects (e.g., people) and edges relations (e.g., friendship) among them.  As we will show, the mathematical characterization of the ``object'' map brings in both new challenges and opportunities. On one hand, we have to face research questions such as: what is a good map? How to compute efficiently maps? Is it useful to define an algebra for maps? On the other hand, maps can be given a ``machine-readable'' (e.g., in RDF) representation and then can be shared, exchanged, reused and composed.
	%%%
	%%%
	%%%
	\subsection{Maps as mathematical objects}
%
	%
	%%%
	%%%
	The formal notion of map of a graph that we are going to introduce captures the standard map representation and allows for the definition of an algebra for combining maps. The idea of a map $M$ of a graph $G$ is essentially that of representing in a concise way information in terms of connectivity among pairs of distinguished nodes of $G$. By making a parallel with geography, $G$ represents the ``region'' or ``territory'' being abstracted via the map and the distinguished nodes represent ``points'' that are absolutely relevant for the map. Distinguished nodes can be our favourite directors in a graph of directors or scientific papers that are relevant for our research in a graph of bibliographic data.
	We now introduce some basic notation and definitions. Let $G= (V_G,E_G)$ be a directed graph, $V_G$ the set of nodes, $E_G$ the set of edges and $u,v$ nodes in $G$. The notation $u \to v$ denotes an edge $(u,v) \in E_G$ and  $u \too v$ a path from $u$ to $v$ in $G$.
	%%%
	%%%
	\begin{defn}[Map]
	 A {\em map} $M=(V_M,E_M)$ of $G=(V_G,E_G)$ is a graph such that $V_M \subseteq V_G$ and each edge  $(x,y) \in E_M$  implies $x \too y$ in $G$.
	\label{def:map}
	\end{defn}
	%%
	%%%
	\begin{defn}[Complete Map]
		%A {\em complete map} is a map such that $x \too y$ in $G$ implies $x\too y$ in $M$, {\color{red}$\forall x,y \in V_M$}.
	A {\em map} is complete if $x \too y$ in $G$ implies $x\too y$ in $M$, $\forall x,y \in V_M$.
	\label{def:cmap}
	\end{defn}
	%
	%\vspace{-.3cm}
	%
	
	The previous definitions capture some basic form of map defined over the Web, such as bookmarks. With bookmarks, the set of distinguished nodes is the set of pages in the Web graph that have been marked as interesting. Nevertheless, a set of bookmarks does not represent a \textit{complete map} since no information about their connectivity is available. A possible complete map of the region in Fig.~\ref{fig:region} is shown in Map1 in Fig.~\ref{fig:maps-comp}. It includes some direct edges, for instance, between J. Ford and S. Kubrick although not originally present in the region.
%%%

	However, sometimes even completeness is not enough to summarize information via maps. The direct edge in the complete map between J. Ford and S. Kubrick is useful because it summarizes the fact that S. Kubrick can be reached from J. Ford via some node (O. Welles), which does not belong to the map (see Fig.~\ref{fig:region}). Consider now the edge $e_1$ in Map 1 in Fig.~\ref{fig:maps-comp}, between J. Ford and W. Allen. Compared to the previous case, this edge does not serve the same purpose. In fact, the connectivity between J. Ford and W. Allen is still maintained via S. Kubrick and there is no other path in the region going from J. Ford to W. Allen only passing for non distinguished nodes. Therefore, $e_1$ is redundant. Avoiding redundancy is crucial for the purpose of \textit{minimizing} the amount of information necessary to keep connectivity between pairs of distinguished nodes. We need to refine the notion of map. Let $G=(V_G,E_G)$ be a graph and $N\subseteq V_G$ a set of nodes. We write $u {\too}_{N} v$ if and only if there is a path from $u$ to $v$ in $G$ not passing through intermediate nodes in $N$.
	%%%%%
	%
	%%%%%
	\begin{defn}[Good Map] Let $M=(V_M,E_M)$ be a map of $G=(V_G,E_G)$ such that $V_M\subseteq V_G$.
	\begin{enumerate*}
	 \setlength{\itemsep}{0pt}%
	\item $M$ is route-complete iff $x\too_{V_M} y$ in $G$ 
	implies $x \to y$ in $M$, $\forall x,y \in V_M$;
	\item $M$ is non-redundant iff $x\to y$ in $M$ implies $x\too_{V_M} y$  in $G$, $\forall x,y \in V_M$.
	\end{enumerate*}
	%%%
	%%
	A map is good iff it is complete, route-complete and non-redundant.
	\label{def:good-map}
	\end{defn}
	%%%%
%	\vspace{-.7cm}
	%%
	%%%%
%			%%%%
%
%	%%%	
	 Map 2 in Fig.~\ref{fig:maps-comp} shows the good map of the region in Fig.~\ref{fig:region}. Interestingly, Theorem~\ref{prop:computing-maps} shows that a region admits a unique good map. The next lemma lays the foundations for computing good maps.
	\begin{lem}
	\label{good}
	A map $M=(V_M,E_M)$ over $G$ is good iff  $\forall x,y \in V_M$ ($x \to y$ in $M$ $\Leftrightarrow$ 
	    $x {\too}_{V_M} y$ in $G$).
	\end{lem}
	%%
	%%%

As discussed in the Introduction, a flexible map framework should consider different types of maps. Accurate maps are the region themselves. Good maps are an example of maps that include connectivity. We now introduce \textit{k-maps}, a family of good maps, which considers nodes in the region having some properties.
	%%
	%%%
	\begin{defn}[$k$-maps]
	Let $G= (V_G,E_G)$ be a graph. The $k$-map of $G$ is the {good} map generated by the set of distinguished nodes $\{ v \in V_G: f(v) \geq k \}$, where $f: V_G \rightarrow \mathcal{R}$ is a function measuring some property of the nodes the region.
	 \end{defn}
	%%%
	\vspace{-.142cm}
	% %

The function $f$ can be, for instance, a measure of the centrality of nodes (e.g., PageRank) or a popularity measure (e.g., number of incident edges). 
% % % %
	%
	
	\medskip
	\noindent
	\textbf{Computing Good Maps.}
	\label{sec:good-maps}
	Maps capture information in a region (i.e., a graph) given a set of distinguished nodes. This section sketches two algorithms for computing good maps and their complexity.
	\begin{theorem}
	Let $G=(V_G,E_G)$ be a graph. Given  $N \subseteq V,$ there is a unique good map $M$ over $G$ with $V_M\mbox{=}N$. $M$ can be computed in time:
	\begin{enumerate*}
	\item $\mathcal{O}(|V_M|\times(|V_G\setminus V_M|+|E_G|))$ if $G$ is a general graph.
	\item $\mathcal{O}((|V_M|\times |V_G \setminus V_M|) + |E_G|)$ if $G$ is a DAG.
	\end{enumerate*}
	\label{prop:computing-maps}
	\end{theorem}
	\vspace{-.6cm}
	\subsection{Algebra of Maps}
	%\label{sec:algebra}
 We have the following main result, which shows the properties of a family of maps obtained from a graph (i.e, a region).
	%%%
	\begin{theorem}
	Let $G=(V_G,E_G)$ be a graph and $\MM(G)$ the set of all maps over $G$. $M_i= (V_{M_i},E_{M_i})\in \MM(G)$ is a map.
	\begin{enumerate*}
	\item The binary relation $\sqsubseteq$ over $\MM(G)$, defined by $M_1 \sqsubseteq M_2$ iff $V_{M_1} \subseteq V_{M_2}$,
	%   and $E_1 \subseteq E_2^*$ (where $E^*$ is the transitive closure of $E$)
	  is a partial order on $\MM(G)$.
	\item The order $\sqsubseteq$ induces a Boolean algebra 
	$(\MM(G), \sqcup, \sqcap, G, \emptyset)$, where: \\
	$M_1 \sqcup M_2$ is the unique good map of $G$ over $V_{M_1} {\color{Brown}\cup} V_{M_2} $;
	$M_1 \sqcap M_2$ is the unique good map of $G$ over $V_{M_1}  \cap V_{M_2} $.
	\item There is an isomorphism of Boolean algebras from 
	$({\cal P}(V),\cup , \cap, V, \emptyset)$ to $(\MM(G), \sqcup, \sqcap, G, \emptyset)$, given by  $N \mapsto M_N$ (the unique good map of $N$ over $G$). 
	\end{enumerate*}
	\end{theorem}
	%%%
	%%%
	%
	
	Having well defined operations over maps enables to obtain new maps from other maps. The question is if the re-computation of a map can be (partially) avoided. The next results shows this possibility. For a given graph $G=(V_G,E_G)$ and $S\subseteq V_G$, denote by $S^*_{G}$ the transitive closure of $S$ over $G$, i.e., the graph $(S, \{ (x,y): x\too_S y \text{ in } G \})$.
	%%%
	%%%
	\begin{prop} 
	Let $M_1=(V_{M_1},E_{M_1})$,  and  $M_2=(V_{M_2},E_{M_2})$ be good maps over $G$.
	\begin{enumerate*}
	\item $M_1 \sqcap M_2 =
	 (V_{M_1} \cap V_{M_2})^*_{M_1} \cup    (V_{M_1} \cap V_{M_2})^*_{M_2}$ 
	                               %%%%%   \setminus \Delta^*(M_1 \cap M_2).$
	                              
	\item  $E_{M_1 \sqcup M_2} \subseteq E_{M_1} \cup E_{M_2} 
   \cup  \{ (x,y) \in E_G: 
	  x \in V_{M_1}, y \in V_{M_2} \} 
  \cup \\ \cup  \{ (y,x) \in E_G:x \in V_{M_1}, y \in V_{M_2} \} 
  \cup \{ x\too_{V_{M_1} \cup V_{M_2}}y,
	  x \in V_{M_1}, y \in V_{M_2} \}
\cup\\ \cup \{ y\too_{V_{M_1} \cup  V_{M_2}} x,
	  x \in V_{M_1}, y \in V_{M_2} \}$
	\end{enumerate*}
	\end{prop}
	%%%
%
	
	\begin{cor}
	The map $M_1 \sqcap M_2$ can be computed only based on information available in the maps $M_1, M_2$ and in time  
	${\cal O}(|V_{M_1} \cap V_{M_2}|\times(|V_{M_1}|+|E_{M_1}|+|V_{M_2}| + |E_{M_2}|)$.
	Moreover, the approximation to $M_1 \sqcup M_2$ (modulo redundancy) cannot be computed more efficiently than computing the good map over $V_{M_1}\cup  V_{M_2}$ from scratch.
	\end{cor}
	%%%%
	% % % % % %
% % % % % % %
	\section{Regions and Maps on the Web}
	\label{sec:framework}
	\vspace{-.2cm}
	We briefly discuss the problem of how to declaratively specify Web regions and keep information about connectivity among nodes. This need is codified in the following general problem: given a  graph $G=(V_G,E_G)$ and a set of nodes $N \subseteq V_G$,  construct a subgraph (a region) $R=(V',E')$ of $G$ such that $N \subseteq V'$. 

	%%
%	%%
Faloutsos el al. \cite{Faloutsos2004} address a variant of this problem: given an edge-weighted undirected graph, two vertices $s,t$, and an integer $k$,  find a connected subgraph $H$ of size $k$ containing $s,t$ that maximizes a given goodness function. Other approaches based have been proposed to discover groups of persons (e.g.,~\cite{Adibi2004}) or simplify networks (e.g.,~\cite{Zhou2010}). However, these approaches do not provide algebras to manipulate the objects that are produced. Besides, they assume that the whole $G$ is locally available; this hinders their applicability to distributed graphs such as the Web graph. How to formally specify and obtain regions of graphs? Graph navigational languages partially address this issue. 

A navigational language $\cal{L}$ over a graph $G=(V_G,E_G)$ is a set of functions (``queries'') of the form $V_G \to \operatorname{subgraphs}(G)\times  \mathcal{P}(V_G)$ that assign to each node $v$ a subgraph (the visited nodes and edges) plus a set of distinguished nodes (the resources selected). Current navigational languages (e.g., XPath, nSPARQL~\cite{Perez2010}, NautiLOD~\cite{Fionda2012}) enable finding pairs of nodes connected by a sequence of edge labels matching some pattern (or navigational expression) expressed via regular expressions over the alphabet of edge labels. This is not enough for our goal; the semantics of current navigational languages should be enhanced to output subgraphs instead of sets of pairs of nodes.

We defined a general navigational language to deal with subgraphs besides sets of nodes. The language features two different semantics: i) the {\sc{visited}} semantics, which return all the portion (i.e., region) of the Web graph visited when evaluating an expression; ii) the {\sc{succesful}} semantics only considers paths that successfully led to some result. In other words, it discards parts of the region that do not contribute from the seed node to reach nodes that satisfy the expression. 
%For sake of space we do not report here the complete description of the language, which is available in~\cite{corr}.
% %

\medskip
\noindent
\textbf{Putting all together.}
Summarizing all the machineries developed so far, the high level specification for building maps of the Web is:
\vspace{-.175cm}
\begin{enumerate*}
\item \textit{Specify the resources of interest}: We leverage a general navigational language.

\item {\em Build the region $R$  corresponding to this specification:} We enchanced the semantics of our navigational language to return subgraphs besides sets of pairs of nodes.

\item {\em Build a formal map corresponding to the region $R$:} We build maps from regions by using the map framework discussed in Section~\ref{sec:maps}.
\end{enumerate*}
\subsection{The Implemented System}
% % %
%
We implemented the map framework in a tool, which can be downloaded at the address \url{http://mapsforweb.wordpress.com}.
We discuss now a real-world example.
	\begin{exmp} (\textbf{Maps of influence networks and algebra})
Specify two regions that contain people that have influenced or have been influenced up to distance 6 by Stanley Kubrick (SK) or Tim Berners-Lee (TBL). The ending nodes in the regions must be scientists. Compute maps and use the algebra of maps.
	\end{exmp}

\begin{figure*}[!b]
		 		  		 		 \centering
		 		  		 		 \includegraphics[width=.965\textwidth]{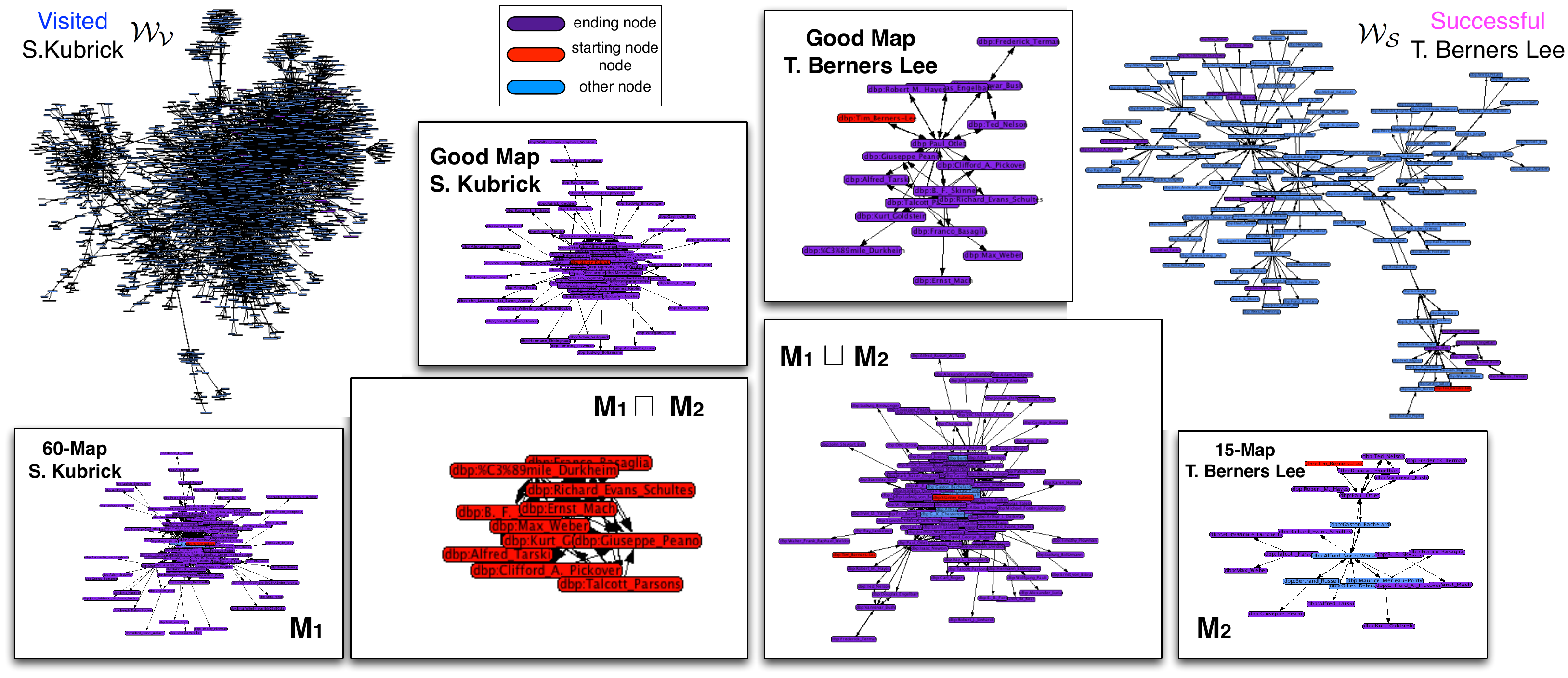}
		 		  		 		  \vspace{-.5cm}
		 		  		 		 \caption{Influence maps of S. Kubrick and T. Berners Lee only considering scientists up to distance 6.}
		 		  		 		 \label{fig:example2}
		 		  		 		 \end{figure*}

	For lack of space Fig.~\ref{fig:example2} only reports regions obtained with the {\sc Visited} semantics. The region associated to the influence network of SK contains 2981 nodes and 7893 edges. The good map associated to SK (109 nodes; 2629 edges) summarizes the region and then provides insight on the connectivity between ending nodes (i.e., scientists that have been influenced or have influenced SK) and with SK. We zoomed in this influence path by computing the 60-map (M$_1$) of the region (120 nodes; 3627 edges).

		 The region associated to TBL is smaller (149 nodes; 236 edges). The associated good map (18 nodes; 43 edges) tells us, for instance, that there exists an influence path from TBL to G. Peano passing via P. Outlet. When zooming in this path, by computing the 15-map (M$_2$) of the region (23 nodes; 43 edges), we discovered that the non ending node B. Russell is also in the path.
		 % % %
		 	%%
		 		  		 		%%% 
		 %
		 	%
		 Fig.~\ref{fig:example2} also shows examples of the algebra of maps. It shows the intersection between M$_1$ and M$_2$. The result is the good map that could have been obtained by making the union of the regions and then computing the good map from the set of distinguished nodes (see Definition~\ref{def:good-map}) given by $V_{M1} \cap V_{M2}$. 
		 
		 However, the advantage of using the algebra is to avoid to compute from scratch the good map and obtain it without looking at the regions. As an example, in the intersection of M$_1$ and M$_2$ we have the nodes G. Peano and A. Tarski, which means that both belong to the influence networks of SK and TBL.
		  The map of the union of M$_1$ and M$_2$ enables to put together information from the two maps; this enables to discover possible additional influence relations between pairs of nodes that are not present in the two maps. In this specific example, there is no path between SK and TBL neither in M$_1$ nor in M$_2$. However, the union of the $k$-maps enabled to discover the connection between TBL and SK (i.e., TBL$\rightarrow$P. Outlet$\rightarrow$B. F. Skinner$\leftarrow$SK).
		  	 % %
		  		 
	%%%	
	\section{Concluding Remarks}
	%%%%
	\label{sec:conclusions}
	%%%
Due to limitations of human I/O capabilities, the management of information at a Web scale calls for automatic mechanisms and thus machine-processable information. In this paper we have shown that maps, key devices in helping human navigation in information spaces, are meaningful on the Web space. 
We think that the formal models presented here are
a starting point for further developing of cartography on the Web. 
% % %
% % %

{\scriptsize
	\bibliographystyle{plain}
	\bibliography{biblio}}
\end{document}